\documentclass[12pt]{article}

\newif\ificonv
\iconvfalse   

\newcommand{\iferm}{\ificonv i \else \fi}
\newcommand{\ifermone}{\ificonv i \else 1\fi}

\def\cala         {{\cal A}}

\def\calb         {{\cal B}}

\def\cald         {{\cal D}}

\def\calf         {{\cal F}}

\newsavebox{\uuunit}
\sbox{\uuunit}
    {\setlength{\unitlength}{0.825em}
     \begin{picture}(0.6,0.7)
        \thinlines
        \put(0,0){\line(1,0){0.5}}
        \put(0.15,0){\line(0,1){0.7}}
        \put(0.35,0){\line(0,1){0.8}}
       \multiput(0.3,0.8)(-0.04,-0.02){12}{\rule{0.5pt}{0.5pt}}
     \end {picture}}
\newcommand {\unity}{\mathord{\!\usebox{\uuunit}}}

\def\ul{\underline}
\def\a{\alpha}
\def\b{\beta}
\def\g{\gamma}
\newif\ifpdf
\ifx\pdfoutput\undefined
   \pdffalse
 \else
   \pdfoutput=1
   \pdftrue
  \usepackage[pdftex]{hyperref}
\fi

\begin{document}
\begin{titlepage}

\font\cmss=cmss10 \font\cmsss=cmss10 at 7pt \leftline{\tt hep-th/0504041}

\vskip -0.5cm \rightline{\small{\tt KUL-TF-05/06}}

\vskip .7 cm

\hfill \vspace{18pt}
\begin{center}
{\Large \textbf{Dirac actions for D-branes \\ on backgrounds with  fluxes}}
\end{center}

\vspace{6pt}
\begin{center}
{\large\textsl{Luca Martucci, Jan Rosseel, \\[2mm] Dieter Van den Bleeken
and Antoine Van Proeyen}}

\vspace{25pt} \textit{Institute for Theoretical Physics, K.U. Leuven,\\ Celestijnenlaan 200D, B-3001 Leuven,
Belgium}\\  \vspace{4pt}
\end{center}

\vspace{12pt}

\begin{center}
\textbf{Abstract}
\end{center}

\vspace{4pt} {\small \noindent The understanding of the fermionic sector
of the worldvolume D-brane dynamics on a general background with fluxes
is crucial in several branches of string theory, like  for example the
study of nonperturbative effects or the construction of realistic models
living on D-branes. In this paper we derive a new simple Dirac-like form
for the bilinear fermionic action  for any D$p$-brane in any supergravity
background, which generalizes the usual Dirac action valid in absence of
fluxes. A nonzero world-volume field strength deforms the usual Dirac
operator in the action to a generalized non-canonical one. We show how
the canonical form can be re-established by a redefinition of the
world-volume geometry.}

\vspace{5cm}

\vfill \vskip 5.mm \hrule width 5.cm \vskip 2.mm {\small \noindent
e-mail: luca.martucci, jan.rosseel, dieter.vandenbleeken,
antoine.vanproeyen@fys.kuleuven.be}

\end{titlepage}

\section{Introduction}

The study of string backgrounds with fluxes turned on has recently received  considerable attention due to its relevance
in the construction of more realistic models  within the framework of string theory. Since D-branes play a crucial role in the
construction of these models, the understanding of the effect of the background fluxes on the world-volume geometry becomes of obvious interest.

While the bosonic sector of the D-brane action on a general background
seems to be completely under control (at least for the Abelian case of a
single D-brane), the introduction of fermions on a general background
required the use of superspace techniques, which allowed to write the
complete  superaction in an elegant and compact form \cite{ced,bt}.
Nevertheless, the use of the superspace as target space hides  how the
background fields enter the fermionic terms of the action and then any
explicit calculation or consideration involving the world-volume fermions
cannot be done only using the implicit superspace formalism. An important
step for the understanding of the fermionic terms in the D$p$-brane
actions in backgrounds with fluxes was obtained in \cite{mms1,mms2}. In
these papers the complete quadratic fermionic action was given for any
D$p$-brane on any (bosonic) supergravity background. The explicit form of
these terms is indeed necessary  in several interesting situations where
the contribution of the world-volume fermionic dynamics becomes relevant.
For example, they are necessary to write the effect of background fluxes
in the effective action governing some phenomenologically interesting
brane configurations \cite{ura1,ura2,louis}.  Also, Euclidean brane
configurations are sources of nonperturbative corrections in lower
dimensional effective theories obtained by compactification
\cite{stro,witten,harvey}. Finally, the knowledge of the explicit form of
the fermionic terms is indeed necessary for any kind of quantum
world-volume computation (see for example \cite{pms}).

In a seminal paper \cite{witten}, Witten found some restrictive
conditions on the possibility of having nonperturbative corrections to
the effective superpotential, which are generally valid for
compactifications on backgrounds without fluxes. Similar kinds of
nonperturbative corrections play a  crucial role  in addressing the
problem of moduli-stabilization in the search for realistic
flux-compactification models along the lines of \cite{kklt}. The
understanding of nonperturbative effects when fluxes are turned on
becomes of obvious importance in this context and this understanding can
pass only through a better understanding of the world-volume fermionic
physics. Indeed some recent work has gone in this direction. For example,
the fermionic quadratic M5-brane action was derived in \cite{sorokin},
also discussing the effect of the background flux in the internal
symmetries of the fermionic action. This action  was then used in
\cite{saulina,kallosh} to discuss new conditions for having
nonperturbative corrections in presence of fluxes. Also, the paper
\cite{trivedi} studies the quadratic fermionic D3-brane action in order
to gain insight into Euclidean D3-brane instantons generating
nonperturbative corrections.

The aim of this paper is to clarify the geometrical structure of the
results derived  in \cite{mms1,mms2}. In the first of these papers
\cite{mms1} the  fermionic bilinear terms in the action of any D$p$-brane
were presented in a  Dirac-like form, making the assumption that the
world-volume field-strength
$\calf_{\alpha\beta}=P[B]_{\alpha\beta}+f_{\alpha\beta}$ was vanishing
($P[.]$ indicates the pull-back on the world-volume while
${f_{(2)}}=dA_{(1)}$ denotes the purely world-volume field strength).
This condition was dropped in the second paper \cite{mms2}, where a
general  $\calf_{(2)}=B_{(2)}+f_{(2)}$ was included. The net effect of
this world-volume field is two-fold. First, it gives a correct
redefinition of the $\kappa$-symmetry operator that naturally enters the
($\kappa$-symmetric) action. Secondly, it adds new kinetic terms that
apparently destroy the Dirac-like form of the actions [see
eq.(\ref{ferm1}) in the next section]. In this paper we will show how
these terms can indeed be reorganized into a more geometrical term that
naturally generalizes the Dirac-like operator. It contains a kinetic term
of the schematic form
\begin{equation}\label{gdirac}
(M^{-1})^{\alpha\beta}\Gamma_\beta\nabla_\alpha\ ,
\end{equation}
where $M_{\alpha\beta}=P[G]_{\alpha\beta}+\calf_{\alpha\beta}$ (see
eq.(\ref{ferm2}) for the precise form). This is somehow analogous to what
happens on the M5 brane with a non-zero $h$ field \cite{sorokin}, and
these two forms should indeed be related by double dimensional reduction.

We also address the problem of writing this kinetic operator in canonical
form. Indeed the presence of a nonzero $\calf_{\alpha\beta}$ enters not
only the kinetic term in the generalized Dirac operator (\ref{gdirac})
but also the generalized integration measure $\sqrt{-\det M}$ that also
appears in the bosonic Dirac-Born-Infeld action. Then it seems natural to
see the effect of the field $\calf_{\alpha\beta}$ as a deformation of the
world-volume geometry. This involves a redefinition of the world-volume
metric and coupling constant analogous to that found in \cite{sw} for
describing the noncommutative theories living on a D-brane in presence of
a constant background $\calf_{\alpha\beta}$ field. Once such a
redefinition is extended to the vielbein, we show how it is possible to
write the bilinear fermionic action in a canonical form.

The paper is structured as follows. In section \ref{sec2} we recall the
results of the papers \cite{mms1,mms2}, namely the explicit form for the
fermionic bilinear action for any D$p$-brane on any (bosonic)
supergravity background, and show how it is possible to rewrite this
result in a new, more geometrical form [see eq.(\ref{ferm2})] that nicely
includes the effect of a nonzero $\calf_{\alpha\beta}$. In section
\ref{sec3} we show how in the new form the action is easily proved to be
invariant in form under T-duality. In section \ref{sec4} we consider the
$\kappa$-fixing of the action and discuss explicitly the possible
resulting world-volume supersymmetries. In section \ref{sec5} we make
some general observations about the natural world-volume geometry which
is deformed by the presence of a non-zero  $\calf_{\alpha\beta}$ and in
section \ref{sec6} we use these observations to show how the fermionic
action can be rewritten in a form containing a canonical Dirac operator
plus mass terms coming from the background fields and the world-volume
configuration itself. Finally, in section \ref{sec7} we present our
conclusions. Appendix \ref{conventions} contains the notation and
comments on the explicit form of the $\kappa $-fixed action are given in
appendix \ref{app:explicit}.

\section{The quadratic fermionic action on a general background}
\label{sec2}

 We are interested in studying the effective actions of D-branes on a general background at the
quadratic order in the fermions. The D-brane actions can be formulated on
a general background using the superspace formalism \cite{ced,bt}.
Unfortunately, even if elegant and in principle complete, this formalism
hides the explicit couplings between the physical fields of the theory
and makes the worldvolume fermionic sector of the theory quite obscure.
An expansion of the superactions in background components is required in
order to have the possibility to make any calculations involving
world-volume fermions. If one starts from the superactions of
\cite{ced,bt} such a calculation can be really cumbersome and requires a
case by case study (see for example \cite{grana} and the recent
\cite{trivedi}). The fermionic quadratic action for any D-brane on any
background was obtained in \cite{mms1,mms2} by following a somehow
different route. The starting point was the normal coordinate expansion
of the M2-brane superaction presented in \cite{grisaru}. Then, by
dimensional reduction and T-duality all the D$p$-brane actions quadratic
in the fermions were derived in a unified and compact form dictated by
the consistency with T-duality.

First of all, let us recall the final result of \cite{mms1,mms2}. The
bosonic part of the D$p$-brane action is given by the standard DBI+CS
form

\begin{equation} \label{bos}
S^{(B)}_{Dp}=-\tau_{Dp}\int d^{p+1}\xi e^{-\Phi}\sqrt{-\det(g+\calf)}+
\tau_{Dp}\int \sum_n P[C_{(n)}]e^{\calf}\ ,
\end{equation}
where $\tau_{Dp}^{-1}=(2\pi)^p (\alpha^\prime)^{\frac{p+1}2}g_s$ is the
brane tension. We use coordinates $\xi^\alpha$, $\alpha=0,\ldots,p$ to
parametrize the worldvolume of the brane,
$g_{\alpha\beta}=P[G]_{\alpha\beta}$ is the pull-back of the background
metric $G_{mn}$ on the world-volume, and
$\calf_{\alpha\beta}=P[B]_{\alpha\beta}+f_{\alpha\beta}$, where
$f_{(2)}=dA_{(1)}$ is the field-strength of the gauge field living on the
brane.

The quadratic fermionic term is given by \cite{mms1,mms2}:
\begin{equation}\label{ferm1}
S^{(F)}_{Dp}=\frac{\iferm\tau_{Dp}}{2}\int d^{p+1}\xi
e^{-\Phi}\sqrt{-\det(g+\calf)}\,\bar\theta(1-\Gamma_{Dp})(\Gamma^\alpha
D_\alpha -\Delta + L_{Dp})\theta\ ,
\end{equation}
where $\Gamma_\alpha$ is the pullback of the gamma matrices $\Gamma _m$
(see for the notation appendix \ref{conventions}, where also some
differences with the conventions used in \cite{mms1,mms2} are spelled
out), the fermionic field $\theta$ is a 10d Majorana spinor in type IIA
and a positive chirality doublet Majorana-Weyl spinor in type IIB. The
other objects entering the actions are the following. For type IIA
D-branes we have
 \begin{eqnarray}
\Gamma_{D(2n)}&=&\sum_{q+r=n}\frac{(-)^{r+1}(\Gamma_{(10)})^{r+1}\epsilon^{\alpha_1\ldots\alpha_{2q}\beta_{1}\ldots\beta_{2r+1}}}{q!(2r+1)!2^q\sqrt{-\det(g+\calf)}}\calf_{\alpha_1\alpha_2}\cdots \calf_{\alpha_{2q-1}\alpha_{2q}}\Gamma_{\beta_{1}\ldots\beta_{2r+1}}\ , \label{chiralA1}\\
L_{D(2n)}&=&\!\!\!\!\!\sum_{q\geq1,
q+r=n}\frac{(-)^{r+1}(\Gamma_{(10)})^{r+1}\epsilon^{\alpha_1\ldots\alpha_{2q}\beta_{1}\ldots\beta_{2r+1}}}{q!(2r+1)!2^q\sqrt{-\det(g+\calf)}}\calf_{\alpha_1\alpha_2}\cdots
\calf_{\alpha_{2q-1}\alpha_{2q}}\Gamma_{\beta_{1}\ldots\beta_{2r+1}}{}^\gamma
D_\gamma\ , \label{LA1}
\end{eqnarray}
whereas for type IIB D-branes
\begin{eqnarray}
\Gamma_{D(2n+1)}&=&\sum_{q+r=n+1}\frac{(-)^{r+1}({\rm i}\sigma_2)(\sigma_3)^r\epsilon^{\alpha_1\ldots\alpha_{2q}\beta_{1}\ldots\beta_{2r}}}{q!(2r)!2^q\sqrt{-\det(g+\calf)}}\calf_{\alpha_1\alpha_2}\cdots \calf_{\alpha_{2q-1}\alpha_{2q}}\Gamma_{\beta_{1}\ldots\beta_{2r}}\ , \label{chiralB1}\\
L_{D(2n+1)}&=&\!\!\!\!\!\!\!\sum_{q\geq 1, q+r=n+1 }\frac{(-)^{r+1}({\rm
i}\sigma_2)(\sigma_3)^r\epsilon^{\alpha_1\ldots\alpha_{2q}\beta_{1}\ldots\beta_{2r}}}{q!(2r)!2^q\sqrt{-\det(g+\calf)}}\calf_{\alpha_1\alpha_2}\cdots
\calf_{\alpha_{2q-1}\alpha_{2q}}\Gamma_{\beta_{1}\ldots\beta_{2r}}{}^\gamma
D_\gamma\ . \label{LB1}
\end{eqnarray}
In the above expressions  $D_m$
(which enters through its pullback $D_\alpha$) and $\Delta$ are the
operators appearing in the supersymmetry transformation laws of the
background gravitino and dilatino and are defined in appendix
\ref{conventions} in the equations (\ref{D}), (\ref{DA}) and (\ref{DB}).

Let us recall that the action (\ref{ferm1}) was first found in
\cite{mms1} in the restricted case in which $\calf_{\alpha\beta}=0$. It
is evident from (\ref{ferm1}) that in this case the action takes an
explicit canonical Dirac-like form, where the background fluxes
contribute with mass terms through the operators $D_\alpha$ and $\Delta$.
The effect of a nonzero $\calf_{\alpha\beta}$ is two-fold. First it is
included in the $\Gamma_{Dp}$ operators (\ref{chiralA1}) and
(\ref{chiralB1}). Secondly, we have the new terms $L_{Dp}$, which enter
the action and which not only add new non-derivative couplings to the
background and worldvolume fields, but also add new kinetic terms. We are
now going to show how these new terms can be reorganized such that the
fermionic lagrangian (\ref{ferm1}) is written in a more geometrical form
where the effect of the field $\calf_{\alpha\beta}$ can be reabsorbed in
a shift of the pulled-back world-volume metric $g_{\alpha\beta}$. This
simplifies the final form of the action considerably and makes its
structure more transparent.

First of all, let us introduce the operator
\begin{equation}\label{gamma0}
\Gamma_{Dp}^{(0)}=\frac{\epsilon^{\alpha_1\ldots\alpha_{p+1}}}{(p+1)!\sqrt{-\det
g}}\Gamma_{\alpha_1\ldots\alpha_{p+1}}.
\end{equation}
It is useful for the manipulations below to notice that it squares to
$(-1)^{(p-1)(p-2)/2}$. We use the formula
\begin{equation}\label{formula1}
\epsilon^{\alpha_1\ldots\alpha_r\beta_{r+1}\ldots\beta_{p+1}}\Gamma_{\beta_{r+1}\ldots\beta_{p+1}}=(-)^{r(r-1)/2}(p+1-r)!
\sqrt{-\det g}\Gamma^{\alpha_1\ldots\alpha_r}\Gamma_{Dp}^{(0)}
\end{equation}
 to write the chiral operators (\ref{chiralA1}) and
(\ref{chiralB1}) in the form used in the paper \cite{bt}
\begin{eqnarray}
\Gamma_{D(2n)}=\frac{\sqrt{-\det g}}{\sqrt{-\det
(g+\calf)}}\Gamma_{D(2n)}^{(0)}(\Gamma_{(10)})^{n+1}\sum_{q}\frac{(-)^q(\Gamma_{(10)})^q}{q!2^q}
 \Gamma^{\alpha_1\ldots\alpha_{2q}}\calf_{\alpha_1\alpha_2}\cdots\calf_{\alpha_{2q-1}\alpha_{2q}}\ , \label{chiralA2}\\
\Gamma_{D(2n+1)}=\frac{\sqrt{-\det g}}{\sqrt{-\det
(g+\calf)}}\Gamma_{D(2n+1)}^{(0)}(\sigma_3)^{n+1}(-{\rm
i}\sigma_2)\sum_{q}\frac{(\sigma_3)^q}{q!2^q}
 \Gamma^{\alpha_1\ldots\alpha_{2q}}\calf_{\alpha_1\alpha_2}\cdots\calf_{\alpha_{2q-1}\alpha_{2q}}\ . \label{chiralB2}
\end{eqnarray}


\noindent {}From (\ref{LA1}) one analogously calculates that
\begin{eqnarray}
L_{D(2n)}&=&\frac{-\sqrt{-\det g}}{\sqrt{-\det (g+\calf)}}\Gamma^{(0)}_{Dp}\left(\Gamma_{(10)}\right)^n\times\nonumber\\
&&\sum_{q\geq1}\frac{\left(-\Gamma_{(10)}\right)^{q-1}}{(q-1)!2^{q-1}}\Gamma^{\a_1\ldots\a_{2q-1}}\calf_{\a_1\a_2}\ldots\calf_{\a_{2q-3}\a_{2q-2}}\calf_{\a_{2q-1}}{}^{\g}D_\g\ .
\end{eqnarray}
This expression can in turn be rewritten as
\begin{eqnarray}
L_{D(2n)}&=&-\Gamma_{D(2n)}\Gamma_{(10)}\Gamma^{\a}\calf_\a{}^\b D_\b - \frac{\sqrt{-\det g}}{\sqrt{-\det (g+\calf)}}\Gamma^{(0)}_{Dp}\left(\Gamma_{(10)}\right)^{n+1}\times\nonumber\\
&&\sum_{q\geq2}\frac{\left(-\Gamma_{(10)}\right)^{q-2}}{(q-2)!2^{q-2}}\Gamma^{\a_1\ldots\a_{2q-3}}\calf_{\a_1\a_2}\ldots\calf_{\a_{2q-5}\a_{2q-4}}\calf_{\a_{2q-3}}{}^{\g_1}\calf_{\g_{1}}{}^{\g_2}D_{\g_2}
\ .
\end{eqnarray}
By iterating this last step (and by doing an analogous calculation for
the IIB case) the following formulae can be found:
\begin{eqnarray}\label{newls} L_{D(2n)}&=& -\Gamma_{D(2n)}\sum_{q\geq
1}(\Gamma_{(10)})^q(\calf^q)^{\alpha\beta}\Gamma_\alpha D_\beta\ ,\cr
L_{D(2n+1)}&=& -\Gamma_{D(2n+1)}\sum_{q\geq
1}(-\sigma_3)^q(\calf^q)^{\alpha\beta}\Gamma_\alpha D_\beta\ ,
\end{eqnarray} where
$(\calf^q)^{\alpha\beta}=\calf^\alpha{}_{\gamma_1}\calf^{\gamma_1}{}_{\gamma_2}\cdots\calf^{\gamma_{q-2}}{}_{\gamma_{q-1}}\calf^{\gamma_{q-1}\beta}$.
We introduce the operator
\begin{equation}\label{hatm}
\tilde M_{\alpha\beta}= g_{\alpha\beta}+\tilde{\Gamma}_{(10)}
\calf_{\alpha\beta}\,,
\end{equation}
where
\begin{equation}
  \mbox{type IIA : }\tilde\Gamma_{(10)}=\Gamma_{(10)}, \qquad
  \mbox{type IIB : }\tilde\Gamma_{(10)}=\Gamma_{(10)}\otimes \sigma_3\ .
  \label{tildeGamma10}
\end{equation}
By putting these results together, it is easy to see that we can write
the fermionic action (\ref{ferm1}) in the compact and elegant form
\begin{equation}\label{ferm2}
S^{(F)}_{Dp}=\frac{\iferm\tau_{Dp}}{2}\int d^{p+1}\xi
e^{-\Phi}\sqrt{-\det(g+\calf)}\,\bar\theta(1-\Gamma_{Dp}) \big[ (\tilde
M^{-1})^{\alpha\beta}\Gamma_\beta D_\alpha -\Delta\big]\theta\ .
\end{equation}

This action represents one of the main result of this paper. The operators $L_{Dp}$ that in the action (\ref{ferm1}) were disturbing,
can now be seen as the source of the natural geometrical coupling to the matrix
\begin{equation}\label{mm}
M_{\alpha\beta}=g_{\alpha\beta}+\calf_{\alpha\beta}\ .
\end{equation}
This matrix substitutes the metric already in the bosonic action or in
the definition of the natural volume element defined on the brane. In
this way we see how the effect of the $\calf_{\alpha\beta}$ field, given
by  the couplings contained in the operators $L_{Dp}$, can be
schematically reabsorbed in the following redefinition of the kinetic
term
\begin{equation}\label{replacegGN}
g^{\alpha\beta}\Gamma_{\beta}\nabla_{\alpha}\quad\rightarrow\quad
 (\tilde M^{-1})^{\alpha\beta}\Gamma_\beta D_\alpha\ .
\end{equation}
On the other hand, the effect of the $\calf_{\alpha\beta}$ field included
in the $\Gamma_{Dp}$ operators appears in the most natural form for a
$\kappa$-symmetric action. This form of the action is related to the
expansion of the super D-brane equations of motion found in
\cite{Bandos:1997rq} for a general background where the structure of
(\ref{ferm2}) and (\ref{replacegGN}) can be recognized.

The effect of a nonzero  $\calf_{\alpha\beta}$ on the world-volume geometry will be considered more carefully in sections \ref{sec5} and \ref{sec6}.
But let us first of all consider two relevant aspects regarding the action (\ref{ferm2}), namely the consistency with T-duality and the $\kappa$-fixed
form of the above action together with its possible linearly and nonlinearly realized supersymmetries. These will be discussed in the following two sections.

\section{Consistency with T-duality}\label{sec3}

The original fermionic action (\ref{ferm1}) was constructed in
\cite{mms2} by using T-duality and assembling  the different terms  in a
rather indirect way,  using the partial results obtained previously in
\cite{mms1} and completing them by means of consistency conditions. Let
us rederive the proof of the T-duality consistency of the above actions
given in \cite{mms1,mms2} starting from their new expression given in
(\ref{ferm2}). This is an important consistency check and it clarifies
also the validity of the arguments, given in \cite{mms1,mms2}, to obtain
the final form of the action (\ref{ferm1}). In order to do this, let us
first of all prove that the term
\begin{equation}\label{tf1}
\frac{\iferm\tau_{Dp}}{2}\int d^{p+1}\xi
e^{-\Phi}\sqrt{-\det(g+\calf)}\,\bar\theta \big[ (\tilde
M^{-1})^{\alpha\beta}\Gamma_\beta D_\alpha -\Delta\big]\theta\ ,
\end{equation}
in the fermionic action (\ref{ferm2}) is left invariant in
form by T-duality. One can check this property directly by using the
usual T-duality rules for the bosonic fields and  Hassan's T-duality
rules for the fermions \cite{hassan}. It is however easier to derive this
property in a less direct way. Let us first  introduce the following
combination of bosonic and fermionic fields
\begin{eqnarray}\label{sfields}
&& {\bf \Phi}= \Phi -\frac{\ifermone}{2}\bar\theta\Delta\theta\ ,\cr &&
{\bf G}_{mn}= G_{mn}-\iferm\bar\theta\Gamma_{(m}D_{n)}\theta\ ,\cr &&
{\bf B}_{mn}=
B_{mn}-\iferm\bar\theta\tilde\Gamma_{(10)}\Gamma_{[m}D_{n]}\theta\ .
\end{eqnarray}
These can be seen as superfields expanded up to the second
order. One of the basic observations of \cite{ms,mms2} is that, using
Hassan's T-duality rules for fermions \cite{hassan}, these  second order
superfields transform (up to second order) in the same way the
corresponding bosonic fields do. The next step is to notice that the term
(\ref{tf1}) can be seen as the second order term arising in the expansion
of a DBI action for the superfields (\ref{sfields})
\begin{equation}\label{superaction}
S_{Dp}=-\tau_{Dp}\int d^{p+1}\xi
e^{-{\bf\Phi}}\sqrt{-\det(P[{\bf G}+{\bf B}]+f)}\ .
\end{equation}
Once we know that the superfields (\ref{sfields}) transform as the
corresponding bosonic fields under T-duality we can immediately conclude
that the action (\ref{superaction}) is left invariant in form by
T-duality and then also its second order term (\ref{tf1}) is invariant by
T-duality.

It is now easy to see that also the second contribution  in the second
order action (\ref{ferm2})
\begin{equation}
-\frac{\iferm\tau_{Dp}}{2}\int d^{p+1}\xi
e^{-\Phi}\sqrt{-\det(g+\calf)}\,\bar\theta\Gamma_{Dp} \big[ (\tilde
M^{-1})^{\alpha\beta}\Gamma_\beta D_\alpha -\Delta\big]\theta\ ,
\end{equation}
is left invariant by T-duality. Since we already know that the term
$\big[ (\tilde M^{-1})^{\alpha\beta}\Gamma_\beta D_\alpha -\Delta\big]$
is invariant in form under T-duality, we need only  that the
$\Gamma_{Dp}$ are transformed into themselves under T-duality. But this
is indeed the case by definition, since in \cite{mms2} these operators
were obtained one from the other by using T-duality. Then the whole
action (\ref{ferm2}) is clearly invariant in form under T-duality.

\section{$\kappa$-fixing and supersymmetry}\label{sec4}
The action (\ref{ferm2}) presented in section \ref{sec2}, once completed
with the bosonic one (\ref{bos}), has world-volume diffeomorphisms and
$\kappa$-symmetry as world-volume gauge symmetries. In this section we
would like to discuss some aspects regarding  their gauge-fixing and the
consequent effects on the way the possible background supersymmetries are
realized on the brane. Let us start by considering the $\kappa$-symmetry.
We recall that, up to the fermionic order we are interested in, the
$\kappa$-symmetry transformations of the bosonic plus  fermionic action
written in (\ref{bos}) and  (\ref{ferm2}) are given by \cite{mms2}
\begin{eqnarray}\label{kappa1}
\delta_\kappa\bar\theta &=& \bar\kappa (1+\Gamma_{Dp})\ ,\cr
\delta_\kappa x^m &=&
-\frac{\ifermone}{2}\delta_\kappa\bar\theta\Gamma^m\theta\ ,\cr
\delta_\kappa A_\alpha &=&
\frac{\ifermone}{2}\delta_\kappa\bar\theta\tilde\Gamma_{(10)}\Gamma_\alpha\theta-
\frac{\ifermone}{2}B_{\alpha m}\delta_\kappa\bar\theta\Gamma^m\theta\ .
\end{eqnarray}
In order to consider the problem of $\kappa$-fixing more clearly, it is
convenient to write (\ref{kappa1}) in a double spinor convention for both
type IIA and type IIB, as introduced in appendix \ref{conventions}. In
this notation, the first two transformation rules of (\ref{kappa1}) can
be written in exactly the same form but with a $\Gamma_{Dp}$ given
by\footnote{Note that, for type IIA, the $\Gamma_{Dp}$ in double spinor
notation is not the same as (\ref{chiralA2}) but is given by
$\Gamma_{Dp}^{\rm double}=\sigma_1\Gamma_{Dp}\sigma_1$ due to the chosen
representation of the charge conjugation matrix. On the other hand,  the
$\sigma _1$ factors in (\ref{GammaIIAd}) have already been extracted from
$\Gamma ^{(0)}_{Dp}$, which is thus considered here as the diagonal
matrix in the extension index.} (for both type IIA and IIB)
 \begin{equation}
\Gamma_{Dp}=(-)^p\Gamma_{Dp}^{(0)}(\sigma_3)^{\frac{p(p+1)}{2}}({\rm
i}\sigma_2)\frac{\sqrt{-\det g}}{\sqrt{-\det
(g+\calf)}}\sum_{q}(\sigma_3)^q\frac{\Gamma^{\alpha_1\ldots\alpha_{2q}}}{q!2^q}
 \calf_{\alpha_1\alpha_2}\cdots\calf_{\alpha_{2q-1}\alpha_{2q}}\ .
 \end{equation}
On the other hand, the last transformation of (\ref{kappa1}) takes the
form
 \begin{equation}
 \label{kappa1b}
\delta_\kappa A_\alpha =
-\frac{\ifermone}{2}\delta_\kappa\bar\theta\sigma_3\Gamma_\alpha\theta-
\frac{\ifermone}{2}B_{\alpha m}\delta_\kappa\bar\theta\Gamma^m\theta\ .
\end{equation}

As stressed in \cite{kalloshD}, it is important to note that the
$\Gamma_{Dp}$ operators entering the $\kappa$-symmetry transformations
(\ref{kappa1}) are off-diagonal:
  \begin{equation}
   \Gamma_{Dp}=\left( \begin{array}{cc} 0 & \check\Gamma_{Dp}^{-1} \\
\check\Gamma_{Dp} & 0 \end{array}\right)\ ,
\end{equation} where
\begin{equation}
\check\Gamma_{Dp}=(-)^{\frac{(p-2)(p-3)}{2}}\Gamma_{Dp}^{(0)}\frac{\sqrt{-\det
g}}{\sqrt{-\det
(g+\calf)}}\sum_{q}\frac{\Gamma^{\alpha_1\ldots\alpha_{2q}}}{q!2^q}
 \calf_{\alpha_1\alpha_2}\cdots\calf_{\alpha_{2q-1}\alpha_{2q}}\ ,
\end{equation}
and $\check \Gamma_{Dp}^{-1}(\calf)=(-)^{\frac{(p-1)(p-2)}{2}}\check
\Gamma_{Dp}(-\calf)$ (remember that $\Gamma_{Dp}^{(0)}$ is defined in
(\ref{gamma0})). Indeed, using this property it is easy to see that the
$\kappa$-symmetry transformation rules can be written in terms of an
irreducible 16-dimensional spinor $\kappa$. We can use for example a
spinor $\kappa$ satisfying the condition
$\tilde\Gamma_{(10)}\kappa=-\kappa$ and rewrite the transformations
(\ref{kappa1}) in the following way
\begin{equation}\label{kappa2}
 \delta_\kappa\bar\theta_1 =\bar\kappa
\check\Gamma_{Dp}\quad ,\quad \delta_\kappa\bar\theta_2 =\bar\kappa\ .
\end{equation}
Then, it is  clear that, as discussed for example in
\cite{bkop,kalloshD}, we can impose a covariant gauge-fixing like
$\tilde\Gamma_{(10)}\theta=\theta$ (i.e. $\theta_2=0$). The resulting
$\kappa$-fixed action can be easily seen to be expressible in terms of
only $\theta_1$ in the following way
\begin{eqnarray}
\label{ferm2b} S^{(F)}_{Dp}&=&\frac{\iferm\tau_{Dp}}{2}\int d^{p+1}\xi
e^{-\Phi}\sqrt{-\det(g+\calf)}\,\Big\{ \bar\theta_1\big[
(M^{-1})^{\alpha\beta}\Gamma_\alpha D^{(0)}_\beta
-\Delta^{(1)}\big]\theta_1 +\cr
&&\quad\quad\quad\quad\quad-\bar\theta_1\check\Gamma_{Dp}^{-1}\big[
(M^{-1})^{\alpha\beta}\Gamma_\beta W_\alpha
-\Delta^{(2)}\big]\theta_1\Big\}\ ,
\end{eqnarray}
where the operators involved are defined in appendix \ref{conventions}.
The explicit form of this action involves terms that vanish by means of
the symmetry properties of the gamma matrices (for example the term
containing the gradient of the dilaton present in $\Delta^{(1)}$ is
identically zero). See appendix \ref{app:explicit} for more on this
point.

We can now pass to the discussion of how possible background
supersymmetries are realized on the world-volume. Let us first of all
recall that, if the supergravity background is supersymmetric, i.e.
possesses a killing spinor $\varepsilon$, then the gauge-unfixed D-brane
action is symmetric under the following (leading order) induced
supersymmetry transformations \cite{mms2} (in standard notation for IIA)
 \begin{eqnarray}\label{wvsusy1}
\delta_\varepsilon\theta &=& \varepsilon \ ,\cr
 \delta_\varepsilon x^m
&=& -\frac{\ifermone}{2}\bar\theta \Gamma^m\varepsilon\ ,\cr
\delta_\varepsilon A_\alpha &=& \frac{\ifermone}{2}\bar\theta
\tilde\Gamma_{(10)}\Gamma_\alpha\varepsilon- \frac{\ifermone}{2}B_{\alpha
m}\bar\theta\Gamma^m\varepsilon\ .
\end{eqnarray}

These transformations have the same form in double spinor notation, up to
the substitution $\tilde\Gamma_{(10)}\rightarrow -\sigma_3$ in the last
line. In order to write these transformations in their gauge-fixed form,
we have to compensate the possible breaking of the $\kappa$-fixing
condition $\theta=\tilde\Gamma_{(10)}\theta$ by some additional
$\kappa$-transformation. For example, we have to add a
$\kappa$-transformation (\ref{kappa2}) with $\kappa=-\varepsilon_2$.
Then, the resulting supersymmetry transformation of the physical
fermionic field $\theta_1$ becomes
\begin{equation}
\delta_\varepsilon\bar\theta_1=\bar\varepsilon_1
-\bar\varepsilon_2\check\Gamma_{Dp}\ .
\end{equation}
It is clear that supersymmetry is preserved by a classical configuration
only if
\begin{equation}\label{unbs}
\bar\varepsilon_1=\bar\varepsilon_2\check\Gamma_{Dp}^{({\rm cl})}\ ,
\end{equation}
which corresponds to the usual condition
$\bar\varepsilon\Gamma_{Dp}^{({\rm cl})}=\bar\varepsilon$ that has to be
satisfied in order for the D-brane to preserve the background
supersymmetry $\varepsilon$. Then, using the relation (\ref{unbs}), it is
possible to write the preserved supersymmetry transformations for the
$\kappa$-fixed action in terms of only $\varepsilon_1$ as follows
 \begin{eqnarray} \label{susyfp}
\delta_{\varepsilon}\bar\theta_1&=&\bar\varepsilon_1(1-\check\Gamma_{Dp}^{({\rm
cl})-1}\check\Gamma_{Dp})\ ,\cr \delta_{\varepsilon}x^{m}&=&
\frac{\ifermone}{2}\bar\varepsilon_1(1+\check\Gamma_{Dp}^{({\rm
cl})-1}\check\Gamma_{Dp})\Gamma^{m}\theta_1\ ,\cr
\delta_{\varepsilon}A_{\alpha}&=&
\frac{\ifermone}{2}\bar\varepsilon_1(1+\check\Gamma_{Dp}^{({\rm
cl})-1}\check\Gamma_{Dp})\Gamma_{\alpha}\theta_1 + \frac{\ifermone}{2}
B_{\alpha m}\bar\varepsilon_1(1+\check\Gamma_{Dp}^{({\rm cl})-1}
\check\Gamma_{Dp})\Gamma^{m}\theta_1\ .
\end{eqnarray}
These expressions contain only the supersymmetry transformations in
lowest order of fermions (without fermion fields for the transformations
of the fermions, and linear in fermions for the transformations of the
bosons). The supersymmetry of the full action needs higher order terms.
However, such transformations are sufficient for determining the
variations of the action linear in fermions. Therefore, they are exact
supersymmetries for the
completely truncated action quadratic  in both bosons and fermions around
some particular classical configuration. To see what they look like in
this linearized approximation, it is convenient to fix the residual gauge
invariance under world-volume diffeomorphisms in order to identify the
physical worldvolume scalar fields. This could be done by adopting the
standard static gauge condition $x^\alpha=\xi^\alpha$, which means that
only the fluctuations $\delta x^{\widehat m}$ ($\widehat m=p+1,\ldots,9$
labels the transverse directions)  are physical, and one has to impose
the condition $\delta x^\alpha =0$. However, since we are working in a
general curved space, this kind of gauge fixing is not the most
geometrical one due to the arbitrariness of the coordinate choice. It is
then  natural to break explicitly the local $SO(1,9)$ Lorentz invariance
of the theory into $SO(1,p)\times SO(9-p)$ and select a class of adapted
co-vielbein $e^{\ul m}=(e^{\ul \alpha},e^{\ul {\widehat m}})$, such that
the pull-back on the brane of the $e^{\ul {\widehat m}}$ is vanishing and
the pulled-back $e^{\ul \alpha}$ form a world-volume vielbein. Now one
can
 consider the fluctuations of the brane as described by a section $\phi^{\ul{\widehat m}}$ of the normal bundle
(i.e. $\phi^{\ul{\widehat m}}=e^{\ul{\widehat m}}_m\delta x^m$). This
means that the natural gauge fixing condition is
\begin{equation}\label{diffix}
e^{\ul{\alpha}}_m\delta x^m=0\,.
\end{equation}
In order to  write the supersymmetry transformations for the completely
gauge-fixed linearized action, we now have to compensate the
transformation (\ref{susyfp}) with a world-volume diffeomorphism
$\delta\xi^{\alpha}(\epsilon )$ defined by the condition
\begin{equation}
\delta\xi^{\alpha}(\epsilon )P[e^{\ul\alpha}]_{\alpha}=
-e^{\ul\alpha}_{m}\delta_{\epsilon}x^m\ .
\end{equation}
Taking into account this compensation and  using the fact that
$\phi^{\ul{\widehat m}}=0$  when evaluated on the classical
configuration, the linearized gauge-fixed supersymmetry transformations
(\ref{susyfp}) become
\begin{eqnarray} \label{susyf}
\delta_{\varepsilon}\theta_1&=&(M^{-1})^{\beta\alpha}(\nabla^{N}_{\alpha}\phi^{\ul{\widehat
m}})\Gamma_\beta\Gamma_{\ul{\widehat m}}\varepsilon_1 +
(M^{-1})^{\beta\alpha}\Gamma_\beta{}^{\gamma}K_{\gamma\alpha}{}^{\widehat{\ul
m}}\phi_{\widehat{\ul m}} +\cr && -\frac12
(M^{-1})^{\alpha\gamma}(M^{-1})^{\beta\delta}(\phi^{\ul{\widehat
m}}H_{\ul{\widehat m}\gamma\delta}+
f_{\gamma\delta})\Gamma_{\alpha\beta}\varepsilon_1\ ,\cr
\delta_{\varepsilon}\phi^{\ul{\widehat m}}&=&
\iferm\bar\varepsilon_1\Gamma^{\ul{\widehat m}}\theta_1\ ,\cr
\delta_{\varepsilon}A_{\alpha}&=&
\iferm\bar\varepsilon_1\Gamma_{\alpha}\theta_1+ \iferm f^{({\rm
cl})}_{\alpha\beta}\bar\varepsilon_1\Gamma^{\beta}\theta_1+ \iferm
B_{\alpha m}\bar\varepsilon_1\Gamma^{m}\theta_1\ ,
\end{eqnarray}
where $\nabla^{N}_{\alpha}=\partial_\alpha +\frac14{\cal
A}_{\alpha}{}^{\ul{\widehat{m}\widehat{n}}}\Gamma_{\ul{\widehat{m}\widehat{n}}}$
indicates the normal bundle covariant derivative with connection
\begin{equation} \label{normalbundleconn}
\cala_\alpha{}^{\ul{ \hat{m}\hat{n}}} = \Omega_\alpha{}^{\ul{
\hat{m}\hat{n}}}\ ,
\end{equation}
where $\Omega_\alpha{}^{\ul{mn}}$ is the pull-back of the spin connection
of the target-space vielbein $e_m^{\widehat{m}}$ and
$K_{\alpha\beta}{}^{\ul{\hat{n}}}$ is the extrinsic curvature of the
world-volume of the brane and is defined by
\begin{equation}\label{extrinsic}
K_{\alpha\beta}{}^{\ul{\hat{n}}}=K_{\beta\alpha}{}^{\ul{\hat{n}}}=e_{\beta}^{\ul{\delta}}
\Omega_{\alpha\ul{\delta}}{}^{\ul{\hat{n}}}\ .
\end{equation}

The derivation of the first of (\ref{susyf}) is straightforward but
tedious, as it involves several rearrangements using gamma matrix
properties along the lines followed to derive (\ref{newls}) from
(\ref{LA1}) and (\ref{LB1}).

\section{The world-volume geometry}\label{sec5}

In section \ref{sec2} we have seen how it is possible to write the
quadratic fermionic action for a D-brane in the form (\ref{ferm2}) which
makes transparent how the background geometry couples to the world-volume
theory. In this section we explore further the geometry characterizing
the theory that lives on the world-volume of the D$p$-brane. In order to
do this we consider from now on the dynamics of the brane around some
classical configuration and use the condition (\ref{diffix}) to fix the
world-volume reparametrization invariance. Furthermore, we will always
consider the (bosonic) fields as evaluated at their classical value and
write the contribution coming from the dynamical fluctuations explicitly.

The first thing that one immediately notices is that the action
(\ref{ferm2}) is not in a canonical form, in the sense that the kinetic
term is contained in
 \begin{equation}\label{kinf}
(\tilde{M}^{-1})^{\alpha\beta}\Gamma_\beta D_\alpha\ ,
 \end{equation}
 which clearly does not give a canonical kinetic term for the fermions.

This feature is already visible  in the bosonic action, as is evident
from the fact that the natural integration measure is given by
 \begin{equation}\label{genvol}
\sqrt{-\det(g+\calf)}\ .
 \end{equation}
 As we want to study the
world-volume physics around a particular background brane configuration
with a nonzero $\calf_{\alpha\beta}$, the general fluctuation $\delta
M_{\alpha\beta}$ of (\ref{mm})
 has a lagrangian of the schematic form
  \begin{equation} \label{bexp}\sqrt{-\det
M}(M^{-1}\cdots M^{-1}\delta M\cdots\delta M)\ .
\end{equation}
This means that not
only the natural volume element is given by $\sqrt{-\det M}$, but also
that the lower indices of the different $(\delta M)_{\alpha\beta}$ are
raised not with a metric but with $(M^{-1})^{\alpha\beta}$, analogously
to what happens in the term (\ref{kinf}) of the fermionic action. The
effect is that the kinetic terms arising from
the expansion of the bosonic action are not in canonical form, as in the fermionic case.

Such a deviation from the canonical form is of course given by the
presence of a nontrivial background $\calf$, which in some sense deforms
the world-volume geometry, generating these noncanonical kinetic terms.
The strictly related situation one immediately thinks of is the case in
which $\calf$ is constant, and the background is flat. It is well known
that in this case the effective world-volume theory is described by a
noncommutative DBI action with zero background $\calf$ field \cite{sw}.
In this case the effect of a constant nonzero $\calf_{\alpha\beta}$ can
be completely reabsorbed in a (non-isotropic) noncommutative deformation
of the world-volume theory. It is also important to remember that the
metric and the coupling constant of the noncommutative theory are not the
same as the background ``commutative" ones but are related to these by
the relations
\begin{eqnarray}\label{ncr}
g_{\alpha\beta}^{({\rm nc})}
&=&g_{\alpha\beta}-\calf_{\alpha\gamma}g^{\gamma\delta}\calf_{\delta\beta}\
,\cr g_s^{({\rm nc})} &=&g_s\sqrt{\frac{\det(g+\calf)}{\det g}}\ .
\end{eqnarray}

We are now going to show how the effect of a nonzero background $\calf_{\alpha\beta}$ can also be reabsorbed in
a non-isotropic deformation of the theory that produces a {\em commutative} theory with canonical kinetic terms
and new coupling terms generated by the presence of the nonzero background $\calf_{\alpha\beta}$. This will
involve a redefinition of the world-volume metric and coupling constant (or better, of the dilaton as seen  by
the brane), analogous to those recalled in (\ref{ncr}) for the noncommutative case.

Since in the following section we will work with fermions, it is
convenient to define from the beginning the deformed theory in terms of
the vielbein instead of in terms of the metric. In particular, since we
will work around some fixed classical configuration, we can restrict to
the class of adapted co-vielbeins $e^{\ul m}=(e^{\ul \alpha},e^{\ul
{\widehat m}})$ such that $P[e^{\ul{\widehat m}}]=0$, introduced in the
previous section. The presence of a world-volume field $\calf$
generically breaks the world-volume $SO(1,p)$ symmetry into
$SO(1,1)\times [SO(2)]^{[(p-1)/2]}$, naturally selecting a subclass of
world-volume vielbeins $e^{\ul \alpha}$ such that\footnote{Here and in
the following we often do not write explicitly the pull-back symbol
$P[.]$ and our notation does not distinguish between the world-volume
vielbein and the $e^{\ul{\alpha}}$ belonging to the target space
vielbein. The resolution of these ambiguities should be clear from the
context.}
 \begin{equation}\label{preferred}
\calf=\tanh \phi_0\ e^{\ul 0}\wedge e^{\ul 1}\
+\sum_{r=1}^{[(p-1)/2]}\tan\phi_r\ e^{\ul{2r}}\wedge e^{\ul {2r+1}}\ .
\end{equation}
This form is clearly invariant under the residual $SO(1,1)\times
[SO(2)]^{[(p-1)/2]}$ symmetry. This decomposition has been used in
\cite{bkop} to write the world-volume chiral operators (\ref{chiralA2})
and (\ref{chiralB2}) (up to a sign) in a nice form. Using the same
approach, we will show that the effect of the field $\calf$ can be
reabsorbed in a non-isotropic deformation of the world-volume metric.

Let us first make a preliminary observation. If we define
\begin{equation}
X_{\ul\alpha}{}^{\ul \beta}=\calf_{\ul\alpha}{}^{\ul \beta}\ ,
\end{equation}
then the action of the matrix $(1+X)$ can be seen as the product of a
rotation $\Lambda\in SO(1,1)\times [SO(2)]^{[(p-1)/2]}$ and an operator
$T$ defined as
\begin{equation}
T=\sqrt{1-X^2}\quad,\quad \Lambda=(1+X)T^{-1}\ .
\end{equation} Note that
$[T,\Lambda]=0$ and  $T$ does not break $SO(1,1)\times
[SO(2)]^{[(p-1)/2]}$. These properties  can be immediately understood by
writing $\Lambda$ and $T$ in our preferred vielbein satisfying
(\ref{preferred}):
 \begin{equation}
T_{\ul \alpha}{}^{\ul \beta} = \left(\begin{array}{ccccc} \frac1{\cosh\phi_0} & 0 & 0 & 0   & \ldots \\ 0 & \frac1{\cosh\phi_0}   &  0 & 0 & \ldots \\
0 & 0 & \frac1{\cos\phi_1} & 0 & \ldots \\ 0 & 0& 0 & \frac1{\cos\phi_1}  & \ldots  \\ \vdots & \vdots & \vdots
& \vdots & \ddots \end{array}\right)
 \end{equation}
 and
 \begin{equation}
\Lambda_{\ul \alpha}{}^{\ul \beta} = \left(\begin{array}{ccccc} \cosh\phi_0 & \sinh\phi_0 & 0 & 0   & \ldots \\ \sinh\phi_0 & \cosh\phi_0   &  0 & 0 & \ldots \\
0 & 0 & \cos\phi_1 & \sin\phi_1 & \ldots \\ 0 & 0& -\sin\phi_1 &
\cos\phi_1  & \ldots  \\ \vdots & \vdots & \vdots & \vdots & \ddots
\end{array}\right)\ .
 \end{equation}
Using the matrix $T$ we can define a new non-isotropically deformed
vielbein
\begin{equation}\label{defv}
\hat e^{\ul \alpha}=e^{\ul\beta}T_{\ul \beta}{}^{\ul \alpha}\ ,
\end{equation}
 and consequently a deformed metric
\begin{equation}\label{defmetric}
\hat g_{\alpha\beta}=\eta_{\ul{\alpha\beta}}\hat e^{\ul\alpha}_\alpha\hat
e^{\ul\beta}_\beta\ .
\end{equation}
It is immediate to see that the definition of the deformed metric
(\ref{defmetric}) is completely identical to the ``noncommutative" one
defined in (\ref{ncr}), i.e.
\begin{equation}
\hat g_{\alpha\beta}=
g_{\alpha\beta}-\calf_{\alpha\gamma}g^{\gamma\delta}\calf_{\delta\beta}\
.\end{equation}
Note that this world-volume metric coincides, up to an
overall constant factor, also  with the on-shell metric found in
\cite{hull} from a Polyakov-like action for D-branes.

We also expect a redefinition of the effective world-volume coupling
constant analogous to the noncommutative one presented in (\ref{ncr}).
Since in this case we allow a non-constant $\calf_{\alpha\beta}$, we
expect here a rescaling of the coupling to the background dilaton.
Indeed, this effect can be easily seen writing the natural volume element
entering the action in the following form
\begin{equation}
\sqrt{-\det (g+\calf)}=\sqrt{-\det g\det(1+X)}=\frac{\sqrt{-\det{\hat
g}}}{\sqrt{\det(1+X)}}\ .
\end{equation}
The most straightforward interpretation of this result, guided by the
noncommutative case (\ref{ncr}), is that we have a world-volume theory
defined in a deformed string frame with a standard volume element
$\sqrt{-\det \hat g}$, and a coupling to a world-volume rescaled dilaton
$\widehat\Phi$ defined by
\begin{equation}
e^{\widehat\Phi}=e^\Phi\sqrt{\det(1+X)}\ .
\end{equation}

Let us finally turn to the original motivation of this deformation, that
is the need to obtain a canonical kinetic term. This effect will become
evident in the following section where we consider the fermionic action,
but already in the bosonic action it is possible to see such an effect by
noting that the $(M^{-1})^{\alpha\beta}$ entering the general expansion
(\ref{bexp}) takes the form
\begin{equation}\label{split}
(M^{-1})^{\alpha\beta}=\hat g^{\alpha\beta}-\hat \calf^{\alpha\beta}\ ,
\end{equation}
where $\hat g^{\alpha\beta}$ is the inverse of $\hat g_{\alpha\beta}$ and
$\hat \calf^{\alpha\beta}=\hat e^\alpha_{\ul \alpha}\hat
e^\beta_{\ul\beta} X^{\ul{\alpha\beta}}$. Then we see how in the deformed
theory the inverse metric $\hat g^{\alpha\beta}$ separates completely
from the contribution given by $\hat \calf^{\alpha\beta}$ which can be
directly identified with the ``deformed" version of the background
world-volume field-strength. Then, when formulated in terms of the new
deformed geometry, the kinetic terms come in a canonical form. Obviously,
as the parent noncommutative formulation suggests, we cannot expect that
the effect of a non-zero background $\calf_{\alpha\beta}$ can be
completely reabsorbed in a deformation of the  metric. Indeed the $\hat
\calf^{\alpha\beta}$ appearing in (\ref{split}) adds other couplings in
the expansion (\ref{bexp}), involving also derivatives of the bosonic
world-volume fields, but since $\hat \calf^{\alpha\beta}$ is
antisymmetric, these are different in nature from kinetic terms and can
be interpreted as generalized electromagnetic couplings. In the following
section we will see how the deformation of the world-volume geometry
introduced here will allow us to isolate a kinetic term, writing the
fermionic action as a standard Dirac action plus mass terms coming from
the embedding in a curved background with fluxes and from the
world-volume background  field strength $\calf_{\alpha\beta}$.

\section{A canonical fermionic action}
\label{sec6}

In this section we reconsider the fermionic action (\ref{ferm2})
discussed in section \ref{sec2}, and rewrite it in terms of the deformed
vielbein (\ref{defv}) defined in the previous section.  Let us start by
first noticing that, following \cite{bkop}, it is possible to write our
chiral operators (\ref{chiralA1}) and (\ref{chiralA2}) that enter the
fermionic action in the form
\begin{equation}
\Gamma_{Dp}=e^{R\tilde\Gamma_{(10)}}\Gamma_{Dp}^{(0)\prime}e^{-R\tilde\Gamma_{(10)}}\
. \end{equation}
We have defined the operators
\begin{equation}\label{chiral01}
\Gamma_{Dp}^{(0)\prime}=\left\{  \begin{array}{l} \Gamma_{Dp}^{(0)}(\Gamma_{(10)})^{\frac{p+2}{2}}\quad\quad {\rm for\ type\ IIA}\ ,\\
 \Gamma_{Dp}^{(0)}({\rm i}\sigma_2)(\sigma_3)^{\frac{p+1}{2}} \quad\quad\quad {\rm for\ type\ IIB}\ ,\end{array}\right.
\end{equation}
 and
\begin{equation}
R=\frac14 Y_{\ul{\alpha \beta}}\Gamma^{\ul{\alpha\beta}}
\end{equation}
is a Lorentz generator expressed easily in our preferred vielbein in the
following way:
 \begin{equation}
 Y_{(2)}=\phi_0 e^{\ul 0}\wedge e^{\ul
1}+\sum_{r=1}^{[(p-1)/2]}\phi_r e^{\ul 2r}\wedge e^{\ul {2r+1}}\ .
 \end{equation}

Let us observe  that $R\tilde\Gamma_{(10)}$ generates on each irreducible
component of a type IIA or IIB spinor a Lorentz transformation belonging
to the unbroken $SO(1,1)\times [SO(2)]^{[(p-1)/2]}$, but also that it
rotates the two irreducible components of a type IIA/IIB spinor in
opposite directions. Then, we can now define, for both type IIA and IIB,
a new ``rotated" fermionic field (recalling that the two irreducible
components are rotated in opposite directions)\footnote{A similar
rotation of the fermions accompanied by a vielbein redefinition of the
kind given in (\ref{defv}) was discussed in
\cite{Bandos:1997rq,Akulov:1998bq}.}
\begin{equation}
\Theta={e^{R\tilde\Gamma_{(10)}}\theta} .
\end{equation}
This sort of generalized chiral rotation is naturally accompanied by the
following redefinition of the operators entering the fermionic action
\begin{eqnarray}\label{rop1} \hat D_\alpha^{(0)} &=&
e^{R\tilde\Gamma_{(10)}}D^{(0)}_\alpha e^{-R\tilde\Gamma_{(10)}}\ ,\cr
\hat W_\alpha&=&e^{-R\tilde\Gamma_{(10)}}W_\alpha
e^{-R\tilde\Gamma_{(10)}}\ ,\cr \hat\Delta^{(1)} &=&
e^{-R\tilde\Gamma_{(10)}}\Delta^{(1)} e^{-R\tilde\Gamma_{(10)}}\ ,\cr
\hat\Delta^{(2)} &=& e^{R\tilde\Gamma_{(10)}}\Delta^{(2)}
e^{-R\tilde\Gamma_{(10)}}\ ,
 \end{eqnarray}
where the operators involved are defined in appendix \ref{conventions}.

In the above redefinition, it can be useful to write the operator $W_m$
entering the action (see appendix \ref{conventions} for its explicit
definition), in terms of a new operator $W$ defined by the relation
$W_m=W\Gamma_m$ and then
\begin{equation}\label{rop2}
\hat W=e^{-R\tilde\Gamma_{(10)}}W e^{R\tilde\Gamma_{(10)}}\
.\end{equation} Then it is possible to write the fermionic action
(\ref{ferm2}) in a canonical form. Indeed, using the fact that
\begin{equation}\label{rot}
e^{-R}\Gamma_{\ul
\alpha}e^R=\Lambda_{\ul{\alpha}}{}^{\ul{\beta}}\Gamma_{\ul\beta}
\end{equation} and that for example
\begin{equation}
\label{id}\Lambda_{\ul \alpha}{}^{\ul
\beta}e_{\ul{\beta}}^m=(1+X)_{\ul\alpha}{}^{\ul\beta}\hat e_{\ul
\beta}^m\ ,\end{equation}
 it is possible to show that the action
(\ref{ferm2}) can be written in the form
\begin{eqnarray}\label{ferm3}
S^{(F)}_{Dp}&=&\frac{\iferm\tau_{Dp}}{2}\int d^{p+1}\xi
e^{-\widehat\Phi}\sqrt{-\det\hat
g}\,\Big\{\bar\Theta(1-\Gamma_{Dp}^{(0)\prime}) \big( \hat\Gamma^\alpha
\hat D^{(0)}_\alpha -\hat\Delta^{(1)}\big)\Theta +\cr
&&+\bar\Theta(1-\Gamma_{Dp}^{(0)\prime})e^{-2R\tilde\Gamma_{(10)}} \big(
\hat
g^{\alpha\gamma}[(1+\tilde\Gamma_{(10)}X)^{-1}]_\gamma{}^{\beta}\hat\Gamma_\beta\hat
W \hat\Gamma_\alpha -\hat\Delta^{(2)}\big)\Theta\Big\}\ ,
\end{eqnarray}
where
\begin{equation} \hat\Gamma_\alpha= \hat
e^{\ul\beta}_\alpha\Gamma_{\ul\beta}\ , \qquad \hat\Gamma^\alpha=\hat
g^{\alpha\beta}\hat\Gamma_\beta\ .
\end{equation}

Using the equations (\ref{rot}) and (\ref{id}), it is possible to write
the explicit form of the new operators defined in (\ref{rop1}) and
(\ref{rop2}). To do this, let us first of all extend the operator
$X_{\ul\alpha}{}^{\ul\beta}$ to an operator acting on all the indices, by
simply putting all the remaining components equal to zero. This
redefinition can then be extended in an obvious way to all the other
operators constructed from $X_{\ul\alpha}{}^{\ul\beta}$. Then, for
example, we can define the complete deformed vielbein $\hat e^{\ul m}_m$,
by simply generalizing the definition (\ref{defv}) into the definition
$\hat e^{\ul n}_m T_{\ul n}{}^{\ul m}$. Of course, these kind of extended
redefinitions do only make sense when restricted to the world-volume of
the brane. The operators $\Delta^{(1)}$ and $\Delta^{(2)}$ are defined in
terms of operators ${\cal T}$ of the form (see appendix \ref{conventions}
for the explicit expression)
\begin{equation} {\cal T}\sim {\cal
T}_{m_1m_2\ldots}\Gamma^{m_1 m_2\ldots}\ .
\end{equation}
Then it is possible to see that the corresponding ``hatted" operators
have the form \footnote{Since $\hat{X}_{\a}{}^{\b} =
\hat{e}_{\a}{}^{\ul{\a}} \hat{e}_{\ul{\b}}{}^{\b} X_{\ul{\a}}{}^{\ul{\b}}
= e_{\a}{}^{\ul{\a}} e_{\ul{\b}}{}^{\b} X_{\ul{\a}}{}^{\ul{\b}} =
X_{\a}{}^{\b}$, the objects in (\ref{ferm3}) and (\ref{hatT}) are
unambiguously defined.}
 \begin{equation} \label{hatT} \hat{\cal T}
\sim {\cal T}_{k_1k_2\ldots}\hat\Gamma^{m_1 m_2\ldots}(1 +
\tilde{\Gamma}_{(10)} X)_{m_1}{}^{k_1}(1+\tilde{\Gamma}_{(10)}
X)_{m_2}{}^{k_2}\cdots \ .
 \end{equation}
 One can do a completely analogous
computation for $\hat{W}$, where however in this case
$\tilde{\Gamma}_{(10)}$ in (\ref{hatT}) is replaced by
$-\tilde{\Gamma}_{(10)}$. Note further that
 \begin{equation} \label{etwoR}
e^{-2R\tilde\Gamma_{(10)}}=\frac{1}{\sqrt{1+X}}\sum_q
\frac{(-)^q(\tilde\Gamma_{(10)})^q}{q!2^q}X_{\ul{\alpha_1\alpha_2}}\cdots
X_{\ul{\alpha_{2q-1}\alpha_{2q}}}\Gamma^{\ul{\alpha_1\ldots\alpha_{2q}}}
 \end{equation}
and that the operator $\Gamma_{Dp}^{(0)}$ entering the definition of
$\Gamma_{Dp}^{(0)\prime}$ in (\ref{chiral01}) can be written in terms of
the deformed quantities by simply adding ``hats" everywhere in the
definition (\ref{gamma0}).

It remains to rewrite $\hat{\Gamma}^\alpha \hat D^{(0)}_\alpha$ in terms
of the deformed variables. The contribution by the $B$-field deforms
analogous to (\ref{hatT}). The pull-back of the target space covariant
derivative can be rewritten as:
\begin{eqnarray} \label{splitconn}
\hat{\Gamma}^\alpha \hat{\nabla}_\alpha & = & \hat{\Gamma}^\alpha
\hat{\cald}_\alpha +
\frac{1}{4}\hat{\Gamma}^\alpha\hat{\calb}_{\alpha}{}^{\ul{\alpha\beta}}
\Gamma_{\ul{\alpha\beta}} - \frac{1}{2} X_{\eta}{}^{\beta}
K_{\alpha\beta}{}^{\hat{\ul{n}}}\,\tilde{\Gamma}_{(10)}
\hat{\Gamma}^{\alpha \eta}{}_{\ul{\hat{n}}} + \nonumber \\ & &
\frac{1}{2} \hat{g}^{\alpha\beta}K_{\alpha\beta}{}^{\hat{\ul{n}}}\,
\Gamma_{\ul{\hat{n}}} + \frac{1}{4} \hat{\Gamma}^{\alpha}
\cala_{\alpha}{}^{\ul{\hat{n} \hat{m}}} \Gamma_{\ul{\hat{n} \hat{m}}}\,.
\end{eqnarray}
In this expression we have made use of the splitting of the pull-backed
target space connection into a world-volume connection plus a part
related to the extrinsic curvature and the normal bundle connection,
introduced in (\ref{normalbundleconn}) and (\ref{extrinsic}). The normal
bundle connection and extrinsic curvature are given by the embedding in
the target space, and as such they do not depend on the world-volume
geometry. As we want to find the kinetic term in a canonical form,
 we needed to introduce in (\ref{splitconn}) a covariant derivative $\cald$ with respect to the deformed frame,
 since this is the frame in which the world-volume geometry is naturally
 described.
 This new derivative is defined using the connection $\hat{\omega}$ of the deformed vielbein, which is related to the
 original world-volume connection ($\omega_\alpha{}^{\ul{\alpha\beta}}  = \Omega_\alpha{}^{\ul{\alpha\beta}}$) by:
\begin{equation} \label{hatomega}
\omega_{\alpha}{}^{\ul{\alpha\beta}} =
{\hat{\omega}}_\alpha{}^{\ul{\alpha\beta}} +
\hat{\calb}_\alpha{}^{\ul{\alpha\beta}}\,,
\end{equation}
with,
\begin{eqnarray} \label{defB}
\calb_\alpha{}^{\ul{\alpha\beta}}&=&(1+\tilde\Gamma_{(10)}X)^{[\ul{\alpha}|\ul{\rho}}\Big[(1+\tilde\Gamma_{(10)}X)^{-1}\hat{\cald}_{\ul\rho}
X(1+\tilde\Gamma_{(10)}X)^{-1} \nonumber \\ & &
-(1-X^2)\hat{\cald}_{\ul\rho}
X\Big]_{\ul{\g}}{}^{|\ul{\b}]}\,\hat{e}_\alpha^{\ul{\gamma}}
\,\tilde\Gamma_{(10)} -\left[\hat{\cald}_\alpha
X(1+\tilde\Gamma_{(10)}X)^{-1}\right]^{[\ul{\alpha\beta}]}\tilde\Gamma_{(10)}\
,
\end{eqnarray}
where we have raised and lowered the flat indices by using
$\eta_{\ul{\alpha\beta}}$ and its inverse as usual. This discussion makes
explicit that the operator $\hat{\Gamma}^\alpha\hat{D}^{(0)}_\alpha$
contains the covariant derivative with respect to the deformed metric
together with some covariant couplings of the world-volume fields to the
background. If one takes the world-volume geometry as given by the
deformed metric (\ref{defmetric}), it can be seen that (\ref{ferm3})
consists of a canonical Dirac operator together with some additional
interactions, given by the embedding and the fluxes.

Let us next consider the effect of gauge fixing kappa-symmetry. We impose
the gauge fixing condition
\begin{equation} \label{kappafixcond}
\Theta = \tilde{\Gamma}_{(10)} \Theta\,,
\end{equation}
which simply means that the second component of $\Theta$ is equal to
zero. It can then easily be seen that the action (\ref{ferm3}) written in
terms of the first component of $\Theta$ (that we indicate again with
$\Theta$) reduces to:
\begin{eqnarray} \label{fixaction}
S^{\prime (F)}_{Dp}&=&\frac{\iferm\tau_{Dp}}{2}\int d^{p+1}\xi
e^{-\widehat\Phi}\sqrt{-\det\hat g}\,\Big\{\bar\Theta \big(
\hat\Gamma^\alpha \hat D^{(0)}_\alpha -\hat\Delta^{(1)}\big)\Theta \cr
&&-\bar\Theta \check\Gamma_{Dp}^{-1} \big( \hat
g^{\alpha\gamma}[(1+X)^{-1}]_\gamma{}^{\beta}\hat\Gamma_\beta\hat W
\hat\Gamma_\alpha -\hat\Delta^{(2)}\big)\Theta\Big\}\ .
\end{eqnarray}
The explicit couplings defined by this action are obtained in appendix
\ref{app:explicit}.

We conclude this section by writing the linearized supersymmetry
transformation rules (\ref{susyf}) in the new deformed variables:
 \begin{eqnarray}
\label{susyf2}
\delta_{\varepsilon}\Theta&=&\hat\Gamma^{\alpha}\nabla^{N}_\alpha\phi^{\ul{\widehat
m}}\Gamma_{\ul{\widehat m}}\chi
-X_{\alpha}{}^{\beta}K_{\beta\gamma}{}^{\ul{\widehat
m}}\phi_{\ul{\widehat m}}\widehat\Gamma^{\alpha\gamma}\chi
-\frac12(\phi^{\ul{\widehat m}}H_{\ul{\widehat m}\alpha\beta}+
{f}_{\alpha\beta})\hat\Gamma^{\alpha\beta}\chi\ ,\cr
\delta_{\chi}\phi^{\hat{\ul{m}}}&=& \iferm\bar\chi\Gamma^{\ul{\widehat
m}}\Theta\ ,\cr \delta_{\chi}A_{\alpha}&=&
\iferm\bar\chi\hat{\Gamma}_{\alpha}\Theta + \iferm B_{\alpha
m}e^{m}_{\ul{\widehat m} }\bar\chi\Gamma^{\ul{\widehat m}}\Theta\ ,
\end{eqnarray}
where $\chi = e^R \varepsilon_1$ and the scalar fields
$\phi^{\ul{\widehat m}}$ describe the brane fluctuations in the normal
directions and $f_{\alpha\beta}$ is the dynamical world-volume
field-strength. Note that the world-volume part of these transformations
takes the usual form valid when the world-volume field-strength
$\mathcal{F}_{\alpha\beta}$ is vanishing while the terms in
(\ref{susyf2}) involving explicitly $B_{(2)}$ and its field-strength
$H_{(3)}$   are non-zero only if some of the off-diagonal components
$B_{\alpha m}e^{m}_{\ul{\widehat m} }$ of $B_{(2)}$ are nonvanishing.

\section{Conclusions}
\label{sec7}

The main results of this paper are the transparant formulae for the
quadratic fermionic part of the D$p$-brane actions on any supergravity
background including possible fluxes. In particular,  (\ref{ferm2}) gives
this parametrization and $\kappa $-symmetric action, and
(\ref{fixaction}) gives a convenient $\kappa $-fixed form.

We started by re-expressing the results of \cite{mms1,mms2} by
re-organizing terms in a more compact notation. We clarified the
underlying geometric structure, using the tensor $\tilde
M_{\alpha\beta}$, see (\ref{hatm}), in both type IIA and IIB. The
formulation makes the invariance under T-duality easy to be verified.
Using a similar doublet notation for IIA and IIB, the $\kappa $-gauge
fixing can be discussed uniformly. The preserved supersymmetry
transformations after gauge-fixing the $\kappa $-symmetry and worldvolume
reparametrizations are obtained in linearized form. In order to clarify
the world-volume geometry we have identified a new natural world-volume
vielbein such that the measure of integration is its determinant and all
kinetic terms for the fermions are recollected in the standard Dirac
operator. Also the supersymmetry transformations in this new metric are
obtained, and the explicit form of the terms entering the $\kappa $-fixed
action is discussed.

These new results can be useful for any kind of quantum calculation on
the brane, in particular for the understanding of non-perturbative
effects in string theory. The results can also be useful for constructing
effective actions for string configurations where D-branes are involved.

\vskip 2cm

\begin{center}
{\large  {\em Acknowledgments}}
\end{center}
We would like to thank J. Gomis, P. Silva and especially D. Sorokin for
helpful discussions. This work is supported in part by the FWO -
Vlaanderen, project G.0235.05, by the Federal Office for Scientific,
Technical and Cultural Affairs through the ``Interuniversity Attraction
Poles Programme -- Belgian Science Policy" P5/27 and by the European
Community's Human Potential Programme under contract MRTN-CT-2004-005104
`Constituents, fundamental forces and symmetries of the universe'. JR and
DVdB are Aspirant FWO-Vlaanderen.

\newpage

\appendix

\section{Conventions}\label{conventions}
In this paper we use Latin indices $m,n,\ldots=0,\ldots,9$ for
10-dimensional curved coordinates, whereas for D$p$-brane world-volume
coordinates we use Greek indices $\alpha,\beta,\ldots=0,\ldots,p$. The
corresponding flat indices are underlined, e.g. the vielbein is given by
$e^{\ul{m}}=e^{\ul{m}}_n dx^n$. The ten dimensional (flat) gamma matrices
are $\Gamma_{\ul m}$ and the 10-dimensional chiral operator is
$\Gamma_{(10)}=\Gamma^{\ul{01\cdots 9}}$. Pulled back gamma matrices are
then $\Gamma _\alpha =\Gamma_{\ul m}e^{\ul m}_m\partial _\alpha x^m$. The
Levi-Civita symbol $\epsilon ^{\alpha _1\ldots \alpha _{p+1}}$ is a
density, i.e. it takes values $\pm 1$.

Differently from \cite{mms1,mms2}, in this paper we use the standard
convention for the expansion of the forms in components, i.e. a $p$-form
$\chi^{(p)}$ is expanded as
\begin{equation}
\chi_{(p)}=\frac{1}{p!}\chi_{m_1\ldots m_p}dx^{m_1}\wedge \ldots\wedge
dx^{m_p}\ ,
\end{equation}
whereas the authors of \cite{mms1,mms2} used the superspace convention
where the $dx^m$ in the above expression are multiplied in the opposite
order. Furthermore, since we consider branes with a positive CS term, the
RR-gauge fields $C^{(n)}$ are related to those used in \cite{mms1,mms2}
by the substitution
\begin{equation}
C_{m_1\ldots m_n}\rightarrow (-)^{\frac{n(n-1)}{2}}C_{m_1\ldots m_n}\ ,
\end{equation}
in such a way that the associated differential forms in the two
conventions are the same. \ificonv  \else Another difference with these
papers is that we have changed the definition of the charge conjugation
matrix such that we avoid a factor ${\rm i}$ for any barred spinor. I.e.
we take $\bar \theta ={\rm i}\theta ^T\Gamma^{\underline{0}}$. \fi

To write the supergravity actions, let us start by introducing the
generalized RR field-strengths\footnote{We  are using essentially the
same conventions as in \cite{myers}.}
\begin{eqnarray} &&
F_{(1)}=dC_{(0)}\quad, \quad F_{(2)}=dC_{(1)}\quad,\quad
F_{(3)}=dC_{(2)}+C_{(0)}H_{(3)}\ ,\cr && F_{(4)}=dC_{(3)}+ H_{(3)}\wedge
C_{(1)}\quad,\quad F_{(5)}=dC_{(4)}+ H_{(3)}\wedge C_{(2)} \ ,
\end{eqnarray}
where $H_{(3)}=dB_{(2)}$.

The type IIA supergravity action is the following
\begin{eqnarray}
S_{IIA}&=&\frac{1}{2\kappa_{10}^2}\int d^{10}x\sqrt{-G}\Big\{
e^{-2\Phi}\big[ R +4(\partial \Phi)^2-\frac{1}{2\cdot 3!}
(H_{(3)})^2\big] +\cr &&\quad -\frac{1}{2\cdot 2!}
(F_{2})^2-\frac{1}{2\cdot 4!}(F_{4})^2\Big\}
-\frac{1}{4\kappa_{10}^2}\int B_{(2)}\wedge dC_{(3)}\wedge dC_{(3)}\ ,
\end{eqnarray}
whereas the type IIB supergravity action is
\begin{eqnarray}
S_{IIB}&=&\frac{1}{2\kappa_{10}^2}\int d^{10}x\sqrt{-G}\Big\{
e^{-2\Phi}\big[ R +4(\partial \Phi)^2-\frac{1}{2\cdot 3!}
(H_{(3)})^2\big] +\cr &&\quad -\frac{1}{2\cdot 3!}
(F_{3})^2-\frac{1}{4\cdot 5!}(F_{4})^2\Big\}+\cr &&\quad
+\frac{1}{4\kappa_{10}^2}\int dC_{(2)}\wedge H_{(3)}\wedge
(C_{(4)}+\frac12 B_{(2)}\wedge C_{(2)})\ .
\end{eqnarray}
In both actions $2\kappa^2_{10}=(2\pi)^7\alpha^{\prime 4}g_s^2$ and in
type IIB one has to add the selfduality condition $F_{(5)}=\star F_{(5)}$
by hand at the level of the equations of motion.

The supersymmetry transformations for both IIA and type IIB  can be
written in the form
\begin{equation}
\delta_\varepsilon \psi_m = D_m\varepsilon \quad,\quad \delta_\varepsilon
\lambda = \Delta\varepsilon\ ,
\end{equation}
where $\varepsilon$ is a
Majorana spinor for type IIA and a doublet of Majorana-Weyl spinors of
positive chirality for type IIB. It is useful to split the operators
$D_m$ and $\Delta$ in two
\begin{equation}\label{D}
D_m=D^{(0)}_m+W_m\quad,\quad \Delta=\Delta^{(1)}+\Delta^{(2)}\ .
\end{equation}
In type IIA we have
\begin{eqnarray}\label{DA}
D^{(0)}&=& \nabla_m+\frac{1}{4\cdot
2!}H_{mnp}\Gamma^{np}\Gamma_{(10)}\ ,\cr W_m &=&  -\frac18 e^{\Phi}\Big(
\frac12 F_{np}\Gamma^{np}\Gamma_{(10)}
+\frac{1}{4!}F_{npqr}\Gamma^{npqr}\Big)\Gamma_m\ ,\cr \Delta^{(1)}&=&
\frac12\Big(\Gamma^m\partial_m\Phi +\frac{1}{2\cdot
3!}H_{mnp}\Gamma^{mnp}\Gamma_{(10)} \Big)\ ,\cr \Delta^{(2)}&=& \frac18
e^\Phi\Big(\frac{3}{2!}F_{mn}\Gamma^{mn}\Gamma_{(10)}-\frac{1}{4!}F_{mnpq}\Gamma^{mnpq}\Big)\
, \end{eqnarray}
while in type IIB
\begin{eqnarray}\label{DB} D^{(0)}&=&
\nabla_m+\frac{1}{4\cdot 2!}H_{mnp}\Gamma^{np}\sigma_3\ ,\cr W_m &=&
\frac18 e^{\Phi}\Big[ F_{n}\Gamma^{n}({\rm i}\sigma_2)
+\frac{1}{3!}F_{npq}\Gamma^{npq}\sigma_1 +\frac{1}{2\cdot
5!}F_{npqrt}\Gamma^{npqrt}({\rm i}\sigma_2)\Big]\Gamma_m\ ,\cr
\Delta^{(1)}&=& \frac12\Big(\Gamma^m\partial_m\Phi +\frac{1}{2\cdot
3!}H_{mnp}\Gamma^{mnp}\sigma_3 \Big)\ ,\cr \Delta^{(2)}&=& -\frac12
e^\Phi\Big[F_{m}\Gamma^{m}({\rm i}\sigma_2)+\frac{1}{2\cdot
3!}F_{mnp}\Gamma^{mnp}\sigma_1\Big]\ ,
\end{eqnarray}
where $\nabla_m=\partial_m+\frac14
\Omega_m{}^{\underline{np}}\Gamma_{\underline{np}}$ is the covariant
derivative.

Finally we end this appendix with some comments on the double spinor
notation used from section \ref{sec4} onwards. From the start, the
spinors in type IIB are doublets. This means that $\theta $ stands for
the 64-component spinor
\begin{equation}
  \pmatrix{\theta _1\cr \theta _2}\ ,
 \label{doubletspinor}
\end{equation}
whose 32-component parts are both left-handed, i.e.
$\theta_i=\Gamma_{(10)}\theta_i$ with $i=1,2$. The $\Gamma $ matrices do
not mix with the extension index, i.e. $\Gamma _m\theta $ stands for
\begin{equation}
  \pmatrix{\Gamma _m\theta _1\cr \Gamma _m\theta _2}\ ,
 \label{gammathetaIIB}
\end{equation}
of which both components are now right-handed. In other words, the
Clifford matrices in the large space act as $\Gamma _m\otimes \unity _2$.
The conjugate spinor $\bar \theta $ is represented by
\begin{equation}
 \pmatrix{\bar \theta _1&\bar \theta _2},
 \label{barthetaIIB}
\end{equation}
and is from the right projected onto itself by
$\frac12(1-\Gamma_{(10)})\otimes \unity _2$.

For type IIA, in sections \ref{sec2} and \ref{sec3} and in the
heavily-used references \cite{mms1,mms2} as in many other papers, the two
spinors are combined in a 32-component Majorana spinor $\theta =\theta
_1+\theta _2$, where $\theta_1=\Gamma_{(10)}\theta_1$ (left-handed) and
$\theta_2=-\Gamma_{(10)}\theta_2$ (right-handed). We now define here also
the doublet spinor (\ref{doubletspinor}), where now both 32-component
parts have opposite chiralities. To obtain formulae that are similar to
the IIB formulae, we use in the 64-component representation different
Clifford representations. The Clifford matrices in the large space are
represented by
\begin{equation}
  \Gamma _m^{\rm double}=\pmatrix{0&\Gamma _m\cr \Gamma _m&0}=\Gamma _m\otimes \sigma
  _1\ .
 \label{GammaIIAd}
\end{equation}
The charge conjugation matrix in the large space is taken to be ${\cal
C}\otimes \sigma _1$, where ${\cal C}$ is the $32\times 32$ charge
conjugation matrix. This implies that the conjugate spinor is
\begin{equation}
 \pmatrix{\bar \theta _2&\bar \theta _1}.
 \label{barthetaIIA}
\end{equation}
These two choices imply e.g. that the expression $\delta _\kappa \bar
\theta \Gamma ^m\theta $ maintains its form when we go from the
32-component notation to the 64-component notation. The matrix $\Gamma
_{(10)}$ is still represented by $\Gamma_{(10)}\otimes \unity _2$, but on
the doublet (\ref{doubletspinor}) it acts as $\unity _{32}\otimes \sigma
_3$. Therefore  $\tilde \Gamma_{(10)}$, see (\ref{tildeGamma10}), is
represented on $\theta $ in both IIA and IIB as $\unity _{32}\otimes
\sigma _3$. In any case it anticommutes with the representations of the
$\Gamma $-matrices.

\section{Explicit form of $\kappa $-fixed action}
\label{app:explicit}

As we observed in section \ref{sec4}, the $\kappa$-fixed action
(\ref{ferm2b}) contains terms that simplify thanks to the symmetry
properties of the gamma matrices, and for example the term containing
the gradient of the dilaton disappears. Such simplifications become,
however, more visible in the formulation with the deformed variables
given in (\ref{fixaction}) of section \ref{sec6}, and so we discuss
explicitly this case. The results obtained can be directly applied to the
action (\ref{ferm2b}) when $\calf_{(2)}=0$, since in this limit the two
actions coincide.

One can extract the explicit couplings included in (\ref{fixaction}) in
the following way. The first two terms in the $\kappa$-fixed action can
be simplified  by making a standard analysis based on the symmetry
properties of $\Gamma$-matrices. For example the coupling to the trace of
the extrinsic curvature and that to the gradient of the dilaton in
$\hat{\Delta}^{(1)}$ drop. The argument for the couplings contained in
$\hat{\Delta}^{(2)}$ is a bit more subtle; one will encounter terms of
the form
\begin{equation} \label{couplings}
\bar{\Theta} \Gamma^{(0)}_{D_p} e^{- 2R} F_{m_1 \cdots m_n}
\hat{\Gamma}^{k_1 \cdots k_n} (1+X)_{k_1}{}^{m_1}\cdots
(1+X)_{k_n}{}^{m_n} \Theta\ .
\end{equation}
Using symmetry properties of the $\Gamma$-matrices, it can be shown that
those are equal to
\begin{equation} \label{couplingstrans}
(-1)^{\frac{(p-n)}{2} + a + 1}\, \bar{\Theta} \Gamma^{(0)}_{D_p} e^{ 2R}
F_{m_1 \cdots m_n} \hat{\Gamma}^{k_1 \cdots k_n}
(1-X)_{k_1}{}^{m_1}\cdots (1-X)_{k_n}{}^{m_n} \Theta\ .
\end{equation}
The sign in front depends on $p$, on $n$ (which is the number of $\hat{\Gamma}$-matrices that are contracted with the fluxes) and on the number $a$ of world-volume indices of the fluxes. 
One notices that (\ref{couplingstrans}) can be obtained from (\ref{couplings}) by replacing $X$ with $-X$, up to a possible overall sign. 
Suppose first that the values of $p$, $n$ and $a$ are such that the sign in front of (\ref{couplingstrans}) is equal to $+1$. 
After expanding $e^{\pm 2R}$ along the lines of (\ref{etwoR}), one can
compare (\ref{couplings}) and (\ref{couplingstrans}) at equal order in
this expansion of $e^{\pm 2R}$. If one looks at an odd order in this
expansion, it is easy to see that only the terms with an odd power of $X$
that are contracted with the $\hat{\Gamma}^{k_1 \cdots k_n}$ are
non-vanishing, while for an even order in the expansion of $e^{\pm 2R}$,
the terms with an even power of $X$ that are contracted with
$\hat{\Gamma}^{k_1 \cdots k_n}$ survive. So, in the end, one sees that
only terms with an even power of $X$ (now also counting the ones in
$e^{\pm 2R}$) survive. If the sign in (\ref{couplingstrans}) is equal to
$-1$ a similar reasoning shows that only the terms with an odd power of
$X$ will survive. For the couplings contained in $W$, one first splits
the matrix $\hat g^{\alpha\gamma}[(1+X)^{-1}]_\gamma{}^{\beta}$ into its
symmetric and its antisymmetric part which, once expanded, contain an
even and odd power of $X$'s respectively. One can then do a reasoning
similar as before on both parts. In this, one has to take into account
that for the antisymmetric part there is an extra minus sign in front of
the analogue of (\ref{couplingstrans}). Indeed, the analogues of
(\ref{couplings}) and (\ref{couplingstrans}) are now:
\begin{eqnarray} \label{analoguescouplings}
& & \bar{\Theta} \Gamma^{(0)}_{D_p} e^{- 2R} F_{m_1 \cdots m_n} \hat{\Gamma}_\beta \hat{\Gamma}^{k_1 \cdots k_n} \hat{\Gamma}_\alpha (1+X)_{k_1}{}^{m_1}\cdots (1+X)_{k_n}{}^{m_n} \Theta\ , \nonumber \\
& & (-1)^{\frac{(p-n)}{2} + a + 1}\, \bar{\Theta} \Gamma^{(0)}_{D_p}
e^{2R} F_{m_1 \cdots m_n} \hat{\Gamma}_\alpha  \hat{\Gamma}^{k_1 \cdots
k_n} \hat{\Gamma}_\beta (1-X)_{k_1}{}^{m_1}\cdots (1-X)_{k_n}{}^{m_n}
\Theta\ ,
\end{eqnarray}
where the $\a$- and $\b$-indices are contracted with $\hat
g^{\alpha\gamma}[(1+X)^{-1}]_\gamma{}^{\beta}$. In order to compare these
two expressions along the lines just described, one needs to exchange the
order of $\hat{\Gamma}_\alpha$ and $\hat{\Gamma}_\beta$. For the piece
with the antisymmetric part of $\hat
g^{\alpha\gamma}[(1+X)^{-1}]_\gamma{}^{\beta}$ this will give an extra
minus sign. In summary, terms with an overall even (odd) power of $X$
survive only if $\frac{(p-n)}{2} + a + 1$ is even (odd).



\end{document}